\documentstyle[prl,aps,psfig,floats,twocolumn]{revtex}
\begin{document}
\draft

\twocolumn[\hsize\textwidth\columnwidth\hsize\csname
@twocolumnfalse\endcsname

\title{A lattice Boltzmann study of reactive microflows}

\author{A. Gabriellii$^1$, S. Succi$^{2,3*}$, E. Kaxiras$^{3}$}
\address{$^1$INFM, Dipartimento di Fisica, Universit\`a di Roma
``La Sapienza'', P.le A. Moro 2, 00185 - Roma, Italy,\\
e-mail: {\tt andrea@pil.phys.uniroma1.it}}
\address{$^2$CNR, Istituto di Applicazioni Calcolo,
viale Policlinico 137, 00161 - Roma, Italy\\
e-mail: {\tt succi@iac.rm.cnr.it}} 
\address{$^{3*}$Visiting Scholar, Lyman Lab. of Physics, 
Harvard University, Cambridge, USA}
\address{$^3$Lyman Laboratory of Physics,
Harvard University, Cambridge, USA}

\date{\today}
\maketitle

\begin{abstract}
The role of geometrical micro-barriers
on the conversion efficiency of reactive flows in narrow three-dimensional
channels of millimetric size is investigated. 
Using a Lattice-Boltzmann-Lax-Wendroff code, we show that 
micro-barriers have an appreciable 
effect on the effective reaction efficiency of the device. 
If extrapolated to macroscopic scales, these effects can
result in a sizeable increase of the overall reaction efficiency.
\end{abstract}

] \narrowtext

\section{Introduction}	

One of the outstanding frontiers of modern applied physics/mathematics
consists in the formulation of models and numerical
tools for the description of complex phenomena involving multiple
scales in space and time \cite{KAX}.
An important example of complex multiscale phenomena is the dynamics
of reactive flows, a subject of wide interdisciplinary
concern in theoretical and applied science.
The complexity of reactive flow dynamics 
is parametrized by three dimensionless quantities:
the {\it Reynolds number} $Re=UL/\nu$, the {\it Damkohler number} 
$Da=\tau_h/\tau_c$, and the {\it Peclet} number $Pe=UH/D$. 
Here $U$, $L$ and $H$ denote the macroscopic flow speed and 
longitudinal/transversal lengths of the flow respectively, $\nu$ is
the fluid kinematic viscosity and $D$ the pollutant 
molecular diffusivity.
The quantities $\tau_c$ and $\tau_h$ represent typical timescales
of chemical and hydrodynamic phenomena.
High Reynolds numbers are associated with turbulence.
High Damkohler numbers imply that chemistry is much faster
than hydrodynamics, so that reactions are always
in chemical equilibrium and take place in tiny regions.
In the opposite regime the chemistry is slow and always takes place at
local mechanical equilibrium.
Finally, high Peclet numbers imply that the transported species
stick tightly to the fluid carrier.
Varying $Re-Da-Pe$ and considering different device morphologies
meets with an enormous variety of 
chemico-physical behaviours \cite{ORAN}.
In this work we deal with {\it low-Reynolds, fast-reacting flows
with heterogeneus catalysis}. In particular we wish to gain insights
into the role of geometric micro-irregularities on the 
effective absorption rate of tracer species 
at catalytic boundaries. For a detailed study see also \cite{EPJAP}.

\section{Mathematical model of reactive microflow dynamics}

We deal with a {\it quasi-incompressible, isothermal flow} with 
soluted species transported (advect and diffuse)
by the flow and, upon reaching solid walls,
they undergo {\it catalytic chemical reactions}.
The basic equations of fluid motion are:
\begin{eqnarray}
&&\mbox{(i)}\,\partial_t \rho + div \; \rho {\bf u} = 0\;\;\;\mbox{and (ii)}
\,\partial_t \rho {\bf u}\! +\! div \; \rho {\bf u} {\bf u}=\nonumber\\
&&-\nabla P +\! div[  2\mu (\nabla {\bf u} + (\nabla {\bf u})^T)\! 
+ \lambda \; div {\bf u}]
\nonumber
\end{eqnarray}
where $\rho$ is the flow density, ${\bf u}$ the flow speed,
$P=\rho T$ the fluid pressure, $T$ the temperature
and $\mu, \lambda$ are the shear and bulk
dynamic viscosities respectively (for the present case of quasi-incompressible
flow with $div {\bf u} \simeq 0$ the latter can safely be ignored).
Finally,
${\bf u} {\bf u}$ denotes the dyadic tensor $u_{a}u_{b},\;a,b=x,y,z$.

Multispecies transport with chemical reactions is described
by a generalized continuity-diffusion equation for each of $s=1,...,N_s$
species:
\begin{eqnarray}
\label{TRA}
&&\partial_t C_s + div C_s {\bf u} =\\  
&&div [D_s C_T \nabla (C_s/C_T)] + \dot \Omega_s \delta({\bf x}-{\bf x}_w)\,,
\nonumber
\end{eqnarray}
where $C_s$ denotes the mass density of the 
generic $s$-th species, $D_s$ its mass diffusivity, 
$C_T=\sum_s C_s$ the total mass of transported species
and $\dot \Omega_s$ is a chemical reaction term whose
contribution
is non-zero along the reactive surface described by the
coordinate ${\bf x}_w$ ($\delta(x)$ is the usual Dirac delta function).
In the following the subscripts $w$ and $g$ mean ``wall'' 
(solid) and ``gas'' in a contact with the wall respectively.

According to Fick's law, the outgoing (bulk-to-wall) 
diffusive mass flux (molecules per unit surface and time) is given by
(hereafter species index $s$ is omitted for simplicity):
\[
J_{g\rightarrow w}=-D \partial_{\perp} C_g|_{wall}
\]
where $\partial_{\perp}$ denotes the normal-to-wall component of the gradient.
Upon contact with solid walls, the transported species react according
to the following empirical rate equation: 
\begin{equation}
\dot \Omega \equiv \frac{d C_{w}}{dt} = J_{g\rightarrow w}\frac
{\Delta S}{\Delta V} -K_c C_w\,,
\label{omega}
\end{equation}
where $\Delta V$ is the volume element of the reactive wall and
$\Delta S$ is the surface element across which fluid-wall mass transfer takes
place. In our case the ratio $\Delta V/\Delta S$ is simply the thickness of
the reactive wall. $K_c$ is the chemical reaction rate dictating
species consumption once a molecule is absorbed by the wall. 
In the following we will use the common linear assumption
\begin{equation}
J_{g\rightarrow w}\frac{\Delta S}{\Delta V}=K_w (C_g-C_w)
\label{Jg}
\end{equation}
where $K_w$ is the wall-fluid mass transfer rate.  
In practice
each boundary cell can be regarded as a microscopic
chemical reactor sustained by the mass inflow from the fluid.
Chemistry (Eqs.~\ref{omega}, \ref{Jg})
sets a time-scale for the steady-state mass exchange rate.
At steady state we obtain:
\[
C_w = \frac{K_w}{K_w+K_c} C_g\,.
\]
Hence
\[
J_{g\rightarrow w}\frac{\Delta S}{\Delta V}= \frac{C_g}{\tau_w+\tau_c}\,, 
\]
where $\tau_w=1/K_w$ and $\tau_c=1/K_c$.
These expressions show that finite-rate chemistry ($K_c>0$) 
ensures a non-zero steady wall outflux of pollutant.

\section{The computational method}

The flow field is solved by a Lattice Boltzmann Equation (LBE) method 
\cite{MZ,HSB,LBE,LBGK} while the multispecies transport 
and chemical reactions are handled with a variant of the 
Lax-Wendroff method (LW) \cite{JCP}.
The LW scheme represents a numerically convenient choice recently developed 
to address multicomponent fluid transport (and reaction) within a LBE-like
language. 

\subsection{Multiscale considerations}

\begin{figure}[tbp]
\centerline{\psfig{file=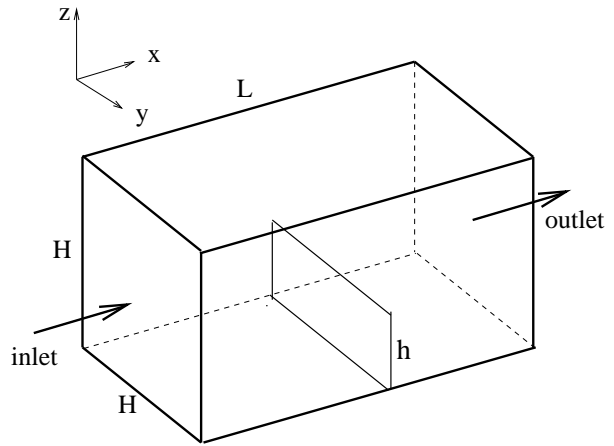,height=6cm,angle=-90}}
\caption{Typical geometrical set up of the channel flow with
a barrier on the bottom wall perpendicular to the flow
of height $h$.}
\label{fig1}
\end{figure}

In this study, the unperturbed 
geometry of the catalytic device is
a straight channel of size
$L$ lattice units along the flow direction (positive $x$ direction)
and $H \times H$ across it ($y$ and $z$ axes).
We add to this unperturbed geometry a
single protrusion (barrier) of unitary thickness
at a fixed $x=L/2$ with  
height $h$ in the $z$ direction, and spanning the channel in 
the $y$ direction (see Fig.\ref{fig1}).

This problem involves at least four relevant time-scales. 
The relevant fluid scales are the advective and 
momentum-diffusive time:
\[
\tau_A = L/U\;\mbox{and}\;
\tau_{\nu}=H^2/\nu\,,
\]
The relevant time-scales for species dynamics are: 
\[
\tau_D=H^2/D,\;\tau_w=K_w^{-1},\,\mbox{and}\;
\tau_c=K_c^{-1}\,.
\]
As discussed in the introduction, they
define the major dimensionless parameters
\begin{equation}
\begin{array}{ll}
Re=UH/\nu \equiv \tau_A/\tau_{\nu},\;
Pe=UH/D \equiv \tau_A/\tau_{D},\\
Da_c=\tau_c/\tau_A,\;\;\; Da_w=\tau_w/\tau_A\,.
\end{array}
\label{Re-t}
\end{equation}
 
\section{Catalytic efficiency}

The device efficiency is defined
as amount of pollutant burned per unit mass injected: 
\begin{equation}
\label{ETA}
\eta = \frac{\Phi_{in}-\Phi_{out}}{\Phi_{in}}\,,
\end{equation}
where $\Phi(x)=\int [uC] (x,y,z) dy dz$
is the longitudinal mass flow of the pollutant at section $x$ and 
$u$ is the $x$ component of ${\bf u}$ ($v$ and $w$ will be the $y$ and $z$ 
components respectively).
The in-out longitudinal flow deficit is equal to the
amount $\Gamma$ of pollutant absorbed at the catalytic wall per unit time.

The goal of the optimization problem is to maximize $\Gamma$ 
at a given $\Phi_{in}$.
This means maximizing
complex configuration-dependent quantities, such as the wall distribution
of the pollutant and its normal-to-wall gradient.
For future purposes, we find it convenient to recast the
catalytic efficiency as $\eta=1-T$, where $T$ 
is the channell {\it transmittance}
$T \equiv \Phi_{out}/\Phi_{in}$.
Roughly speaking, in the limit of
fast-chemistry, this is controlled by the ratio 
of advection to diffusion timescales. 
It is intuitive that high efficiencies are associated with large values
of the ratio $\tau_A/\tau_D$, namely low-Peclet numbers.

\section{The role of micro-irregularities}

We now discuss the main qualitative 
effect of the micro-barrier from a microscopic point of view. 

Firstly, it provides a potential
enhancement of reactivity via the 
increase of the surface/volume ratio.
How much of this potential is actually realized depends
on the resulting flow configuration.

Here, the fluid plays a two-faced role.
First, geometrical restrictions lead to local fluid acceleration,
hence less time for the pollutant to migrate from the
bulk to the wall before being convected away by the mainstream
flow. This effect may become
appreciable on micro-scales 
for micro-flows with $h/H \simeq 0.1$ (like in actual
catalytic converters).
Moreover, obstacles shield away
part of the active surface (wake of the obstacle) where the fluid
circulates at much reduced rates (stagnation) so that
less pollutant is fed into the active surface.
The size of the shielded region is proportional to
the Reynolds number of the flow.
On the other hand, if by some mechanism the flow proves capable
of feeding the shielded region, then efficient absorption is restored
simply because the pollutant is confined by recirculating patterns and has
almost infinite time to react without being convected away.
This case is met mainly in presence of sufficiently energetic turbulent
fluctuations at high values of the {\em micro barrier}-Peclet number
$Pe_h=\frac{w'h}{D} \gg 1$
where $w'$ is the $z$-component of the velocity field at the barrier tip.

With some appropriate approximations \cite{EPJAP},
one can show that the efficency is:
\begin{equation}
\eta_0 \simeq 1-e^{-L/l}\,, 
\label{eta0}
\end{equation}
where $l=l_\perp^2{\bar U}/D$, 
${\bar U}=\sum_{y,z}uC/\sum_{y,z}C$, 
$l_\perp^2=C\tau H^2/(2C_g)$ and $\tau\simeq 
\left(1/\tau_D+1/(\tau_c+\tau_w)\right)^{-1}$.

Note that in the low absorbtion limit 
$L\ll l$, the above relation reduces to $\eta_0 \simeq L/l$, meaning 
that halving, say, the absorption length 
implies same efficiency with a twice shorter catalyzer.
In the opposite high-absorption limit, $L\gg l$, 
the relative pay-off becomes increasingly less significant.

We now turn 
to the case of a ``perturbed'' geometry.
Let us begin by considering a single barrier of height $h$ 
(Fig.~\ref{fig1}).
The reference situation is a smooth channel at high Damkohler
(Eq.~\ref{eta0}).
From \cite{EPJAP} we find an estimate of 
perturbative corrections in the smallness parameter 
$g \equiv h/H$:
\begin{equation}
\label{DETA}
\frac{\delta \eta}{\eta_0}=
\simeq
\frac{A}{2} \frac{h}{H} Re_h [Sc+K\;(a-1)]
\end{equation}
where $A=H/L$ is the aspect ratio of the channel,
$Sc=\nu/D$ is the Schmidt number, and $a$ is a regime dependent 
parameter.
The wake length $W$ can be estimated by 
$W/h = K Re_h$ with $K \simeq 0.1$. 
Three distinctive cases can be identified:
(i) $a=0$: the wake region is totally deactivated, absorption zero;
(ii) $a=1$: absorption in the wake region 
is exactly the same as for unperturbed flow;
(iii) $a>1$: the wake absorption is higher than with unperturbed flow 
(back-flowing micro-vortices can hit the rear side of the barrier).

\section{Application: reactive flow over a microbarrier}

The computational scheme has been applied
to a fluid flowing in a millimeter-sized box of
size $2 \times 1 \times 1$ millimeters along the $x,y,z$ directions
with a perpendicular barrier of height $h$
(see Fig. \ref{fig1}).
Upon using a $80 \times 40 \times 40$ grid, we
obtain a lattice with $dx=dy=dz=0.0025$ ($25\,\mu m$).
We assume a real sound speed $V_s=300$ m/s which becomes 
$c_s=1/\sqrt 3$ in lattice units.
Therefore a time-step is equivalent to
$dt=c_s\; dx/V_s \simeq 50 ns$.

The flow is forced with a constant volumetric force
which mimics the effects of a pressure gradient.
The fluid flow carries a passive pollutant which is continuously injected 
at the inlet with a flat profile across the channel.
Diffusing across the flow, it reaches solid walls where it 
reacts according to  a first order catalytic reaction:
$C+A \rightarrow P$,
where $A$ denote an active catalyzer and $P$ the reaction products.
The initial conditions are:
\[
\begin{array}{lll}
C(x,y,z)=1\;\mbox{at the inlet, and}\;=0\,\mbox{elsewhere}\nonumber\\
\rho(x,y,z)=1\nonumber\\
u(x,y,z) = U_0,\;\;\; v(x,y,z)=w(x,y,z)=0\nonumber\,.
\end{array}
\]
The pollutant is then released at the open outlet, while 
flow periodicity is imposed at the inlet/outlet boundaries.
On the upper and lower walls, the flow speed is forced
to vanish, whereas the fluid-wall mass exchange is
modelled via a mass transfer rate equation of the form
previously discussed.
Our simulations refer to the following values (in lattice units):
$U_0 \simeq 0.1-0.2$, $D=0.1$, $\nu=0.01$, $K_c=K_w=0.1$. 
This implies
$Pe \simeq 40,\;\;\;Re \simeq 400,\;\;\;Da>80$ (see also Eq.~\ref{Re-t}).
In order to study the effects of the barrier 
height $h$, we consider the following
values: $h=0,2,4,8$.
The typical simulation time is $t=32000$ time-steps
(about $1.6$ milliseconds in physical time) corresponding
to two mass diffusion times across the channel.
We may estimate the reference
efficiency for the case of smooth channel:
with $\bar U \simeq 0.1$, and $\tau=20$, we obtain
$l \simeq 200$, hence $\eta_0 \simeq 0.33$. 

\begin{figure}[tbp]
\centerline{\psfig{file=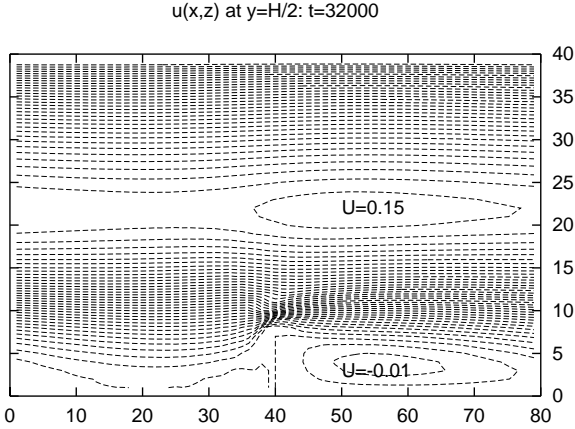,height=6.5cm,angle=-90}}
\caption{Typical two-dimensional cut of the flow pattern with a single barrier
of heigth $h=8$. Streamwise flow speed in the plane $y=H/2$.}
\label{fig2}
\end{figure}

\begin{figure}[tbp]
\centerline{\psfig{file=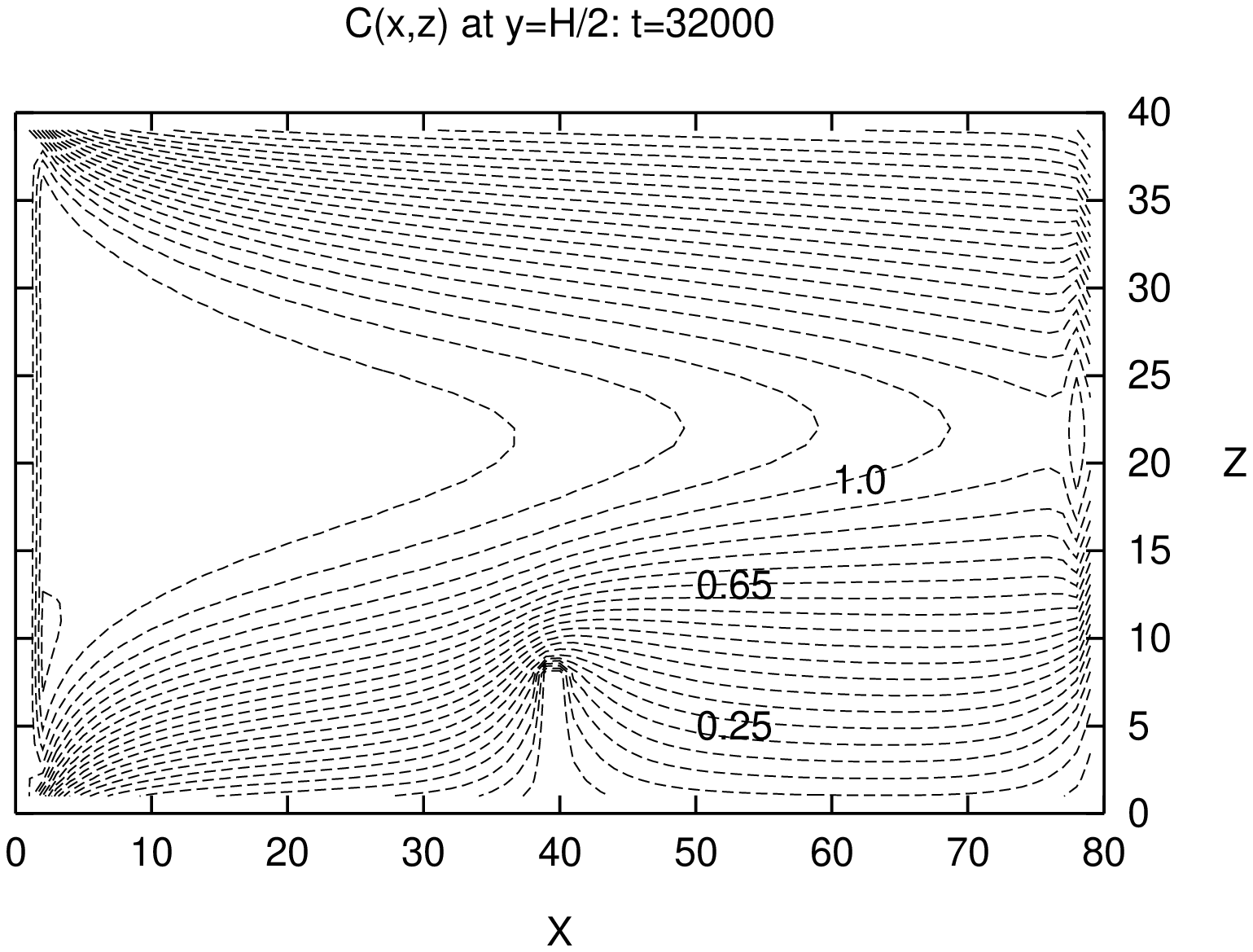,height=6.5cm,angle=-90}}
\caption{Concentration isocontours with a single barrier of heigth $h=8$
on the plane $y=H/2$.}
\label{fig3}
\end{figure}

\begin{figure}[tbp]
\centerline{\psfig{file=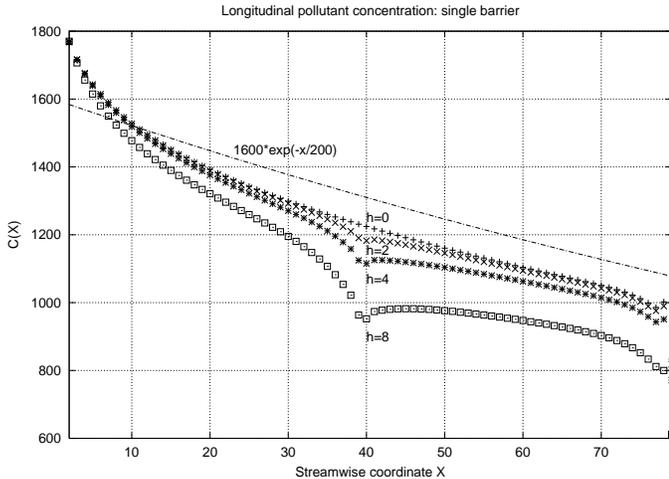,height=6.5cm,angle=-90}}
\caption{Integrated concentration $C(x)$ of the pollutant 
with a single barrier of height $h=0,2,4,8$ after $32000$ steps.
The dashed line represent a theoretical evaluation with no barrier ($h=0$)
and $l\simeq 200$.}
\label{fig4}
\end{figure}

\begin{figure}[tbp]
\centerline{\psfig{file=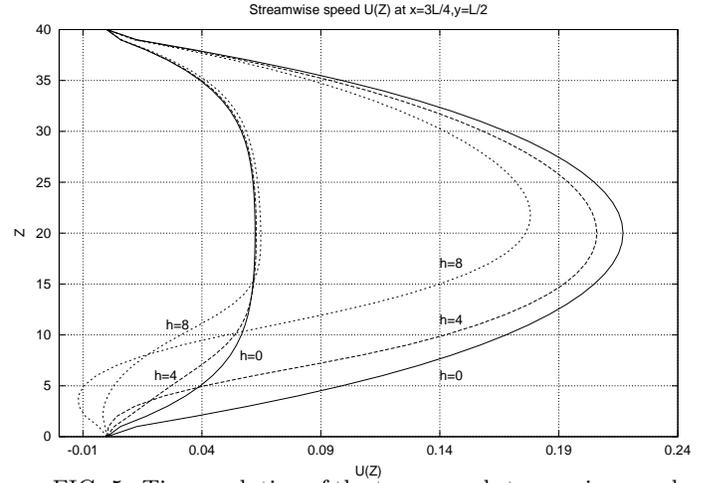,height=6.5cm,angle=-90}}
\caption{Time evolution of the transversal streamwise
speed $u(z)$ at $x=3L/4$ and $y=L/2$. 
Single barrier of varying height $h=0,4,8$ at $t=3200$ and $t=32000$.
Note the backflow for $h=8$ at small $z$.}
\label{fig5}
\end{figure}

\begin{figure}[tbp]
\centerline{\psfig{file=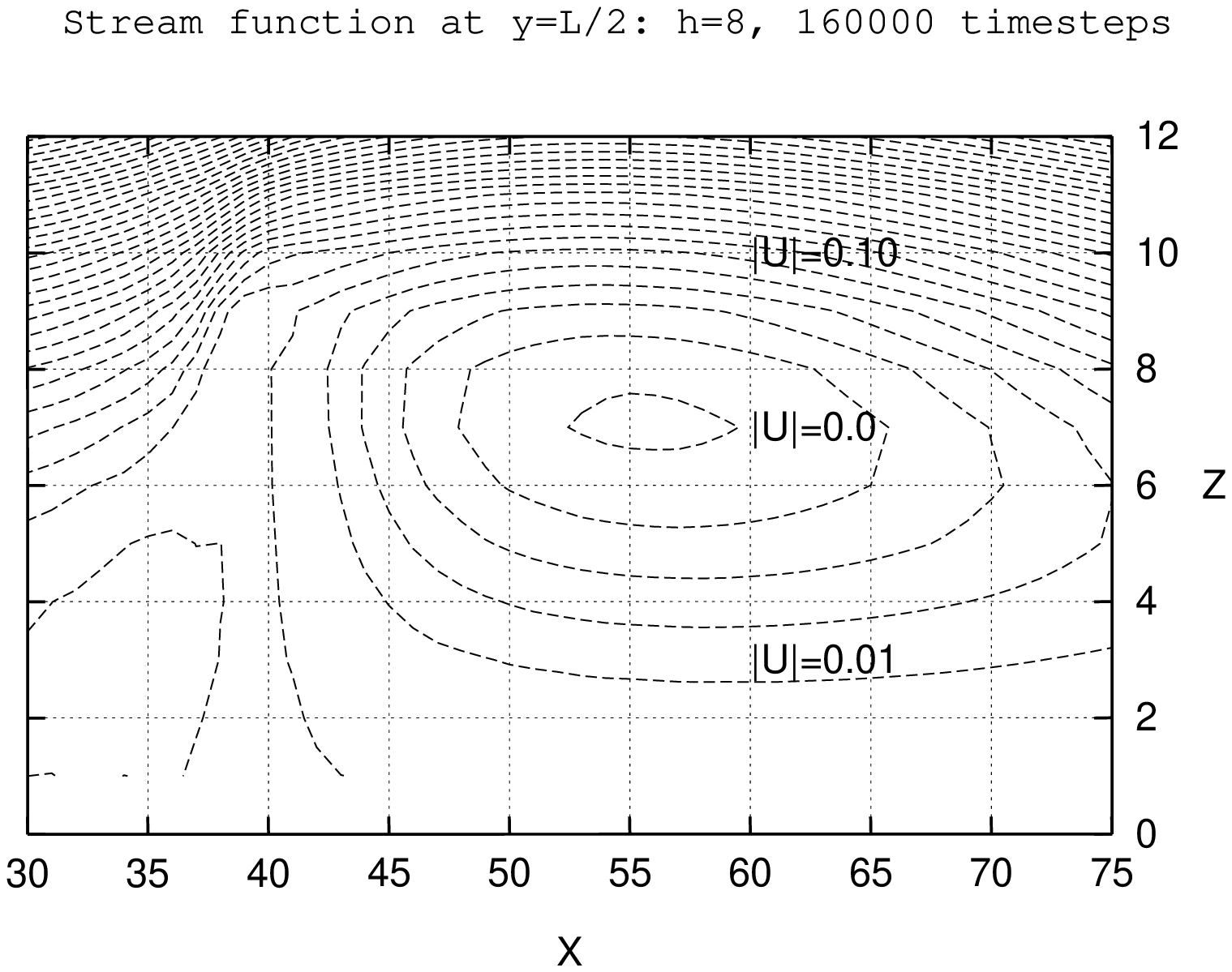,height=6.5cm,angle=-90}}
\caption{Blow-up of the streamlines of the flow field past a 
barrier of height $h=8$ located at $x=40$. The velocity direction
in the closed streamlines of the vortex is clockwise.
The recirculation effects are feeble 
and depletion is dominant.
In fact for $h=8$ the local Peclet number is 
$\sim0.01 \cdot 8/0.1=0.8$, seemingly too small to provide
micro-turbulent effects.}
\label{fig6}
\end{figure}

A typical two-dimensional cut of the flow pattern
and pollutant spatial distribution in the section
$y=H/2$ is shown in Figs.~\ref{fig2} and \ref{fig3}, which refer to the 
case $h=8$.
An extended (if feeble) recirculation pattern is
well visible past the barrier.
Also, enhanced concentration gradients on
the tip of the barrier is easily recognized from Fig.~\ref{fig3}.
The integrated concentration of the pollutant $C(x)=\sum_{y,z} C(x,y,z)$
is presented in Fig.~\ref{fig4} for the cases $h=0,2,4,8$.
The main highlight is a substantial reduction of the pollutant
concentration with increasing barrier height.
We measure also the
the pollutant longitudinal mass flow
$\Phi(x)$. The efficiency $\eta$ is defined
by Eq.~\ref{ETA}. 
The results are shown in Table \ref{tab1}, 
where subscript $A$ refers
to Eq.~\ref{DETA} with $a=1$.
These results are in a reasonable agreement
with the analytical estimate apart deviations $h=8$ for which
the overall efficency is overestimated.
Leaving aside the initial portion of the channel, 
our numerical data are pretty well fitted by an exponential 
with absorption length $l=200$, in a good agreement
with the theoretical estimate $l \simeq 200$.
\begin{table}[tbp]
\begin{center}
\begin{tabular}{|c|c|c|c|c|} \hline
Run  & $h/H$ &   $\eta$  & $\frac{\delta \eta}{\eta},\frac{\delta \eta_A}
{\eta_A}$\\ \hline
R00  &0      &    0.295  & 0.00      \\ \hline
R02  &1/20   &    0.301  & 0.02,0.025\\ \hline
R04  &1/10   &    0.312  & 0.06,0.10 \\ \hline
R08  &2/10   &    0.360  & 0.22,0.40 \\ \hline
\end{tabular}
\end{center}
\caption{Single barrier at $x=40$: the effect of barrier height.}
\label{tab1} 
\end{table}
The barrier also promotes 
a potentially beneficial flow recirculation, which is well visible
in Figs.~\ref{fig5} and \ref{fig6}. 
They clearly reveal a recirculating backflow
for $h=8$. 
For applications to many barriers see \cite{EPJAP}.

\section{Upscaling to macroscopic devices}

\begin{figure}[tbp]
\centerline{\psfig{file=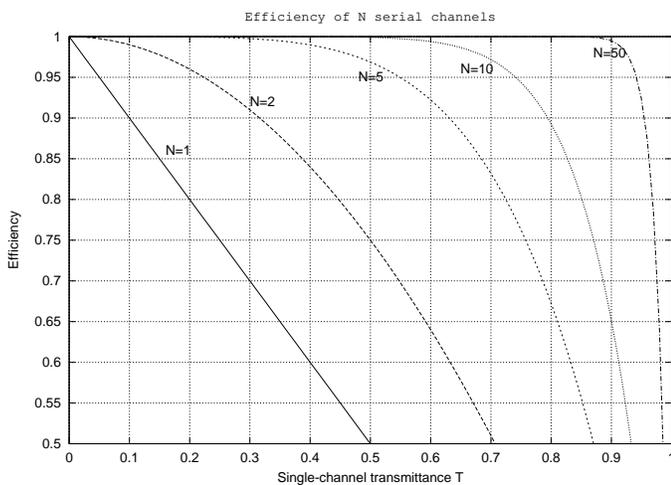,height=6.5cm,angle=-90}}
\caption{Efficiency of a series of $N$ micro-channels as a function
of the single-channel transmittance.}
\label{fig7}
\end{figure}

It is important to realize that even tiny improvements on the
microscopic scale can result in pretty sizeable cumulative
effects on the macroscopic scale of 
the real devices, say $10$ centimeters.
The efficiency of an array
of $N$ serial micro-channels can be estimated simply as
\begin{equation}
\label{BOLD}
\eta_N = 1 -T^N\,.
\end{equation}
It is readily recognized that even low single-channel
efficiencies can result in significant efficiencies
of macroscopic devices with $N=10-100$ (see Fig.~\ref{fig7}). 
Equation \ref{BOLD} with numerical data from present
simulations provide satisfactory agreement with
experimental data \cite{CORRO,SAE}.

Nonetheless, extrapolations based on Eq.~\ref{BOLD}
must be taken very cautiously in the case of rugh or fractal walls
\cite{SAP} or of fully developed turbulence. 

\section{Conclusions}

Although these simulations generally confirm qualitative expectations
on the overall dependence on the major physical parameters, they
also highlight the existence of non-perturbative effects, such as
the onset of micro-vorticity in the wake of geometrical obstrusions,
which are hardly amenable to analytical treatment. 

Work performed under NATO Grant PST.CLG. 976357.

\end{document}